\documentclass{article}
\usepackage{spconf,amsmath,graphicx}
\usepackage{enumitem}
\setlist{nosep, leftmargin=14pt}
\usepackage{mwe} %
\usepackage[dvipsnames]{xcolor}
\usepackage{subcaption}
\usepackage{hyperref}
\usepackage{booktabs}

\title{AI-Assisted Diagnosis for Covid-19 CXR Screening: From Data Collection to Clinical Validation}

\name{\begin{tabular}{c} Carlo Alberto Barbano$^{1,2}$\sthanks{Correspondence: carlo.barbano@unito.it} \quad Riccardo Renzulli$^1$ \quad Marco Grosso$^3$ \quad Domenico Basile$^3$ \\ Marco Busso$^3$ \quad Marco Grangetto$^1$\end{tabular}}
\address{$^1$University of Turin \quad $^2$LTCI, Télécom Paris, IP Paris \quad $^3$Azienda Sanitaria Locale TO3}
\begin{document}
\maketitle
\begin{abstract}
In this paper, we present the major results from the Covid Radiographic imaging System based on AI (Co.R.S.A.) project, which took place in Italy. This project aims to develop a state-of-the-art AI-based system for diagnosing Covid-19 pneumonia from Chest X-ray (CXR) images.  
The contributions of this work are manyfold: the release of the public CORDA dataset, a deep learning pipeline for Covid-19 detection, and the clinical validation of the developed solution by expert radiologists.
The proposed detection model is based on a two-step approach that, paired with state-of-the-art debiasing, provides reliable results. Most importantly, our investigation includes the actual usage of the diagnosis aid tool by radiologists, allowing us to assess the real benefits in terms of accuracy and time efficiency. Project homepage: \url{https://corsa.di.unito.it/}
\end{abstract}

\begin{figure}
    \centering
    \begin{subfigure}[b]{0.485\columnwidth}
        \includegraphics[width=\columnwidth]{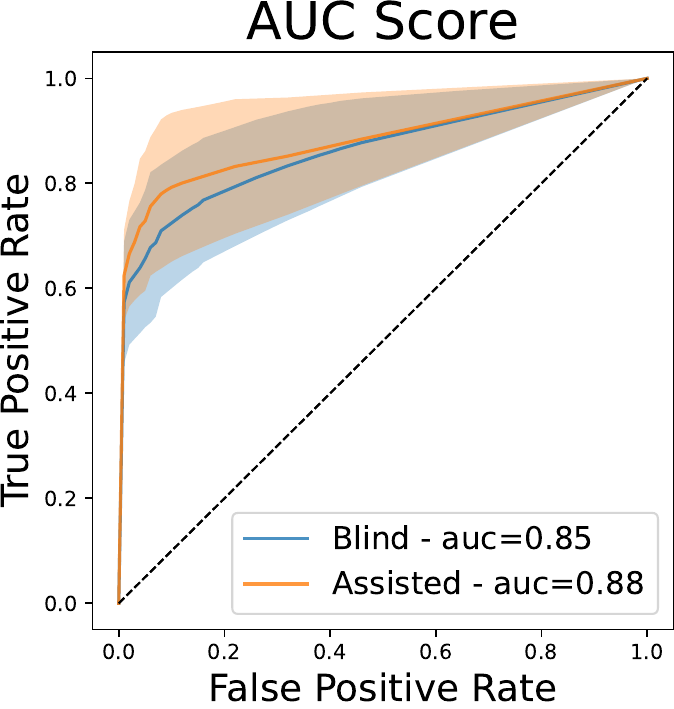}
        \caption{~}
        \label{fig:mean-auc}
    \end{subfigure}
    \begin{subfigure}[b]{0.485\columnwidth}
        \includegraphics[width=\columnwidth]{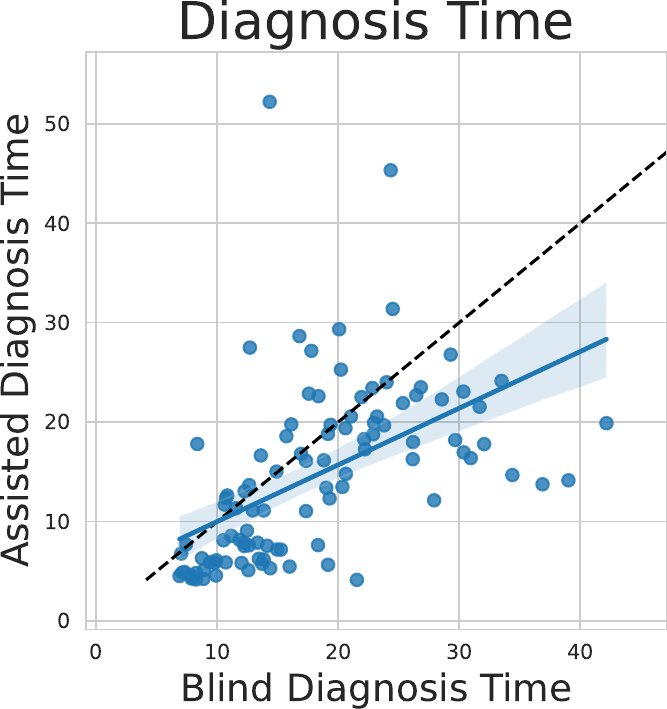}
        \caption{~}
        \label{fig:mean-time}
    \end{subfigure}
    \caption{AI-assisted diagnosis help radiologists in making more accurate and faster dagnosis. In (a) we report the average ROC curve of a pool of radiologists during external validation, in a blind setting (no AI) and in an AI-assisted setting. In (b) we compare the average diagnosis time (per image) in a blind vs. assisted setting. We notice that the average AUC increases in the AI-assisted setting, while the diagnosis is quicker on average.}
    \label{fig:teaser}
\end{figure}

\section{Introduction}
\label{sec:intro}

During the beginning of 2020, Covid-19 rapidly spread in China and out into multiple countries worldwide~\cite{zu2020coronavirus}.  As of November 2023, there were more than 770 million confirmed cases worldwide, with almost 7 million deaths\footnote{\url{https://covid19.who.int/}}.
It is now widely recognized how radiological imaging can contribute to identifying Covid-19 even in the early stages of the disease: Chest X-rays (CXR) and computed tomography (CT) can be helpful diagnostic means. 
Clinical evidence shows that CT represents the gold standard for diagnosing Covid-19~\cite{shi2020radiological}. However, its use in the initial management of the suspected patient is limited by logistical problems (sanitization of the room/suitable dressing/undressing areas of the operators), radioprotection, and, finally, costs. For these reasons, CXR is often chosen in emergency departments as the first radiological method for the ease of execution, the low dose of ionizing radiation, and the more contained costs (if compared to CT).

In 2022 the Piedmont Region funded the Covid Radiographic imaging System based on AI (Co.R.S.A.) project, a collaborative initiative by the University of Turin, two hospital radiological units (A.O.U Citt\'a della Salute e della Scienza, Azienda Sanitaria ASL TO3) and an enterprise (REGOLA),  to develop and validate in the field a state-of-the-art AI-based system for aiding the diagnosis of Covid-19 pneumonia from CXR images. 
In this work, we describe the main project milestones and contributions:

\begin{enumerate}
    \item the release of the CORDA dataset;
    \item a robust deep learning pipeline for Covid-19 detection;
    \item an external clinical validation by expert radiologists.
\end{enumerate}
  
Among the mentioned contributions, the latter is probably the most impactful: indeed, papers on Covid-19 identification from CXR proliferated in recent years, but in most cases a complete validation of the system by radiology doctors is lacking.  
Although the emergency setting of the Covid-19 pandemic peak is fortunately over, the groundwork built by this project can serve as a solid foundation for quick-starting future responses to epidemics, should the need arise.

\section{The CORDA dataset}

CORDA~\cite{corda_dataset} contains 3843 images of different modalities, with 1601 CXR and 2242 CT images\footnote{In the scope of this work, we only considered CXR images.}. The dataset was made publicly available in January 2023\footnote{\url{https://zenodo.org/record/7821611}}. 
This dataset aims to provide a multi-center collection of radiographic images for Covid-19 detection, in order to build more robust machine learning algorithms and models. The dataset curation is part of the ongoing project Co.R.S.A. CORDA comprises four different Italian hospitals:
\begin{enumerate}
    \item A.O.U. Città della Salute e delle Scienza (Molinette), Torino (CDSS);

    \item A.O.U. San Luigi Gonzata, Orbassano (SLG);
    
    \item A.O. Mauriziano, Torino (MRZ);

    \item Centro Cardiologico Monzino, Milano (MNZ). \\
\end{enumerate}
\noindent \textbf{Dataset composition} In Tab.~\ref{tab:corda} we show the distributions of Covid-19 positive and negative CXR cases across the four institutions that contributed to CORDA; the number of images acquired by Computed Radiography (CR) and Digital Radiography (DR) is reported as well. It can be observed that the samples of different institutes are inevitably unbalanced and potentially present different imaging characteristics. As shown in~\cite{glocker2019machine, tartaglione2020unveiling} this can negatively impact deep learning models, limiting their generalization capability. These biases will be considered in Sec.~\ref{sec:method} with proper regularization techniques to limit their effect in the model training phase. 

\begin{table}
\centering
\begin{tabular}{l c c c c}
    \toprule
     \textbf{Institution} & \textbf{Pos. cases} & \textbf{Neg. cases} & \textbf{CR/DR} \\
     \midrule
     CDSS & 362 & 183 & 401/144 \\ 
     SLG & 250 & 477 & 713/14 \\
     MRZ & 138 & 35 & 163/10  \\
     MNZ & 156 & 0 & 63/93 \\
     \midrule
     Total & 906  & 695 & 1340/261 \\
     \bottomrule
\end{tabular}
\caption{CORDA composition: Covid-19 positive and negative patients, imaging modality CR/DR.} \label{tab:corda}
\end{table}
\section{Automatic Diagnosis of Covid-19}

In this section, we present a method for Covid-19 diagnosis based on deep learning (DL).
The method that we propose consists of two steps: first, we pretrain a deep neural network (DNN) on a large-scale CXR dataset, with the aim of detecting objective radiological findings, then we apply transfer learning to train a Covid-19 classifier on the CORDA dataset. As previously shown in~\cite{barbano2022two}, this approach provides reliable results. 
Additionally, we also employ a state-of-the-art debiasing technique, in order to address the presence of possible biases and spurious information related to the acquisition site in the data, a phenomenon commonly known as \emph{site-effect}~\cite{glocker2019machine}.

\subsection{Method}
\label{sec:method}

\begin{table}
    \centering
    \begin{tabular}{l c c c c c}
    \toprule
    & \textbf{CDSS} & \textbf{SLG} & \textbf{MRZ} & \textbf{MNZ} & \textbf{Avg} \\
    \midrule
         Baseline & 0.68 & 0.83 & 0.74 & 0.87 & 0.78 \\
         FairKL  & \textbf{0.70} & \textbf{0.85} & \textbf{0.77} & \textbf{0.88} & \textbf{0.80} \\
    \bottomrule
    \end{tabular}
    \caption{Results of Covid-19 detection on the CORDA dataset, in terms of balanced accuracy.}
    \label{tab:corsa-corda-results-fairkl}
\end{table}

\textbf{Pretraining on objective radiological findings.}
We leverage a large-scale dataset, \emph{CheXpert}~\cite{irvin2019chexpert}, which contains annotation for different kinds of common radiological findings that can be observed in CXR images (like opacity, pleural effusion, cardiomegaly, etc.). This large dataset is well suited for multi-label classification tasks; in fact, more than one finding can be commonly observed simultaneously in ill patients' lungs.
CheXpert provides 14 different types of observations for each image in the dataset. 
We use the setup described in~\cite{barbano2022two}, consisting of a slightly modified DenseNet-121~\cite{huang2017densely} architecture.
\newline
\noindent\textbf{Covid-19 Prediction}
To obtain a Covid-19 classifier, we employ the DNN encoder pre-trained on CheXpert as a frozen feature extractor on the CORDA dataset. We train a fully connected binary classifier for the final prediction, using the standard binary cross-entropy loss (BCE).
\newline
\noindent\textbf{Site regularization}
From a debiasing point of view, we define as bias-aligned all those samples that share the same acquisition site w.r.t. to a given \emph{anchor} sample, and bias-conflicting the ones which do not. Also, we define as \emph{positive} the samples which share the same target class as the anchor. We employ the recently proposed FairKL~\cite{barbano2023unbiased} regularization technique, which aims at minimizing the Kullback-Leibler divergence of the distance distributions in the latent space of positive bias-aligned $B_{+,b}$ and positive bias-conflicting  and $B_{+,b'}$: 
\begin{equation}
    \mathcal{R}^{FairKL} = D_{KL}(B_{+,b} || B_{+,b'})
\end{equation}

This regularization term aims to make samples of the same class indistinguishable in the latent space based on the acquisition site. This choice is motivated by the possible correlation between the target label, modality, and source institution, as shown in Tab.~\ref{tab:corda}. The final objective function we optimize is thus $J = \mathcal{L}^{BCE} + \lambda\mathcal{R}^{FairKL}$, with $\lambda \geq 0$.

\subsection{Results}

We train the final classifier (on the frozen backbone) for 100 epochs, using SGD as optimizer, with a learning rate of 0.01, decayed with a cosine schedule, a batch size of 32 and an image size of 448x448. We also employ an augmentation scheme composed of random crop, rotation, and cutout. For each experiment, we perform a 4-fold cross-validation on the CORDA dataset, and we report the mean results in Tab.~\ref{tab:corsa-corda-results-fairkl}, evaluated separately on each institution test set. We report as \emph{baseline} a simple transfer learning employing only BCE for training. With this approach, we achieve a 78\% balanced accuracy in Covid-19 classification. By employing FairKL (with $\lambda=1$), we achieve a consistent improvement with respect to the baseline, improving the results for each institution and achieving an average balance accuracy of 80\%.

\section{Clinical Validation}

\begin{figure}
    \centering
        \includegraphics[width=\columnwidth]{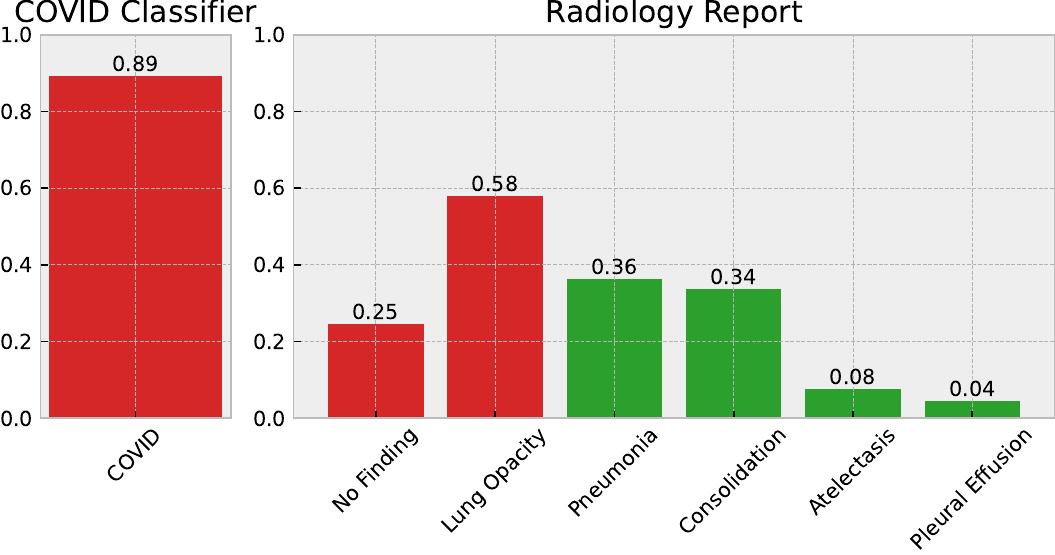}
    \caption{Example of AI report in our custom DICOM viewer. This report is shown to the evaluating radiologist alongside the CXR image. The information includes the predicted probability of Covid-19 infection and other relevant lung pathologies. For easiness of readability, probabilities $> 0.5$ are marked in red (except for the "No Finding" class).}
    \label{fig:dicom-viewer}
\end{figure}

In this section, we focus on assessing the impact of AI-aided diagnosis in the radiologists' workflow, one of the key aspects of our project.
To this end, the clinical validation focuses on measuring two KPIs: \emph{i)} the accuracy of the radiologists' diagnosis \emph{ii.)} the time needed to formulate it.

\subsection{Materials and tools}

\textbf{Validation data} As validation data, 100 external images were collected at ASL To3, of which 50 are from Covid-19 patients and 50 are control cases. These images are not part of the CORDA dataset and were not used for previous phases. This external validation also allows us to assess whether the proposed method can generalize to novel data and sites.
\newline 
\noindent \textbf{Dicom viewer}
For this validation, we developed a custom DICOM image viewer which allows DNNs predictions to be shown alongside the analyzed image. Our software automatically collects the diagnosis made by the radiologist for each image and the time taken to make it. Primary functionalities of a standard DICOM viewer are included, such as image manipulation (e.g. zoom, translation) and adjustable windowing (VOI LUT). Fig.~\ref{fig:dicom-viewer} depicts an example of DNNs predictions as shown to the evaluating radiologist in our viewer.

\begin{figure*}
    \centering
    \begin{subfigure}[b]{\linewidth}
        \includegraphics[width=0.16\linewidth]{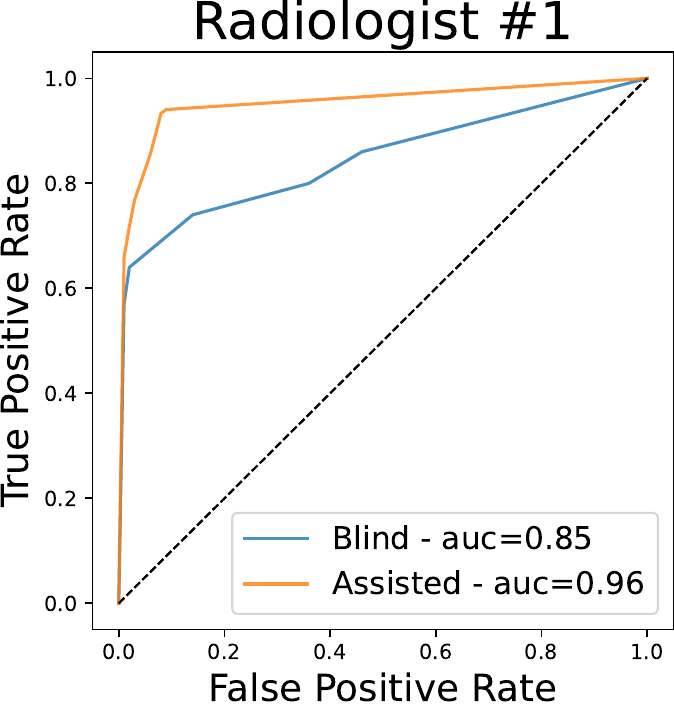}
        \includegraphics[width=0.16\linewidth]{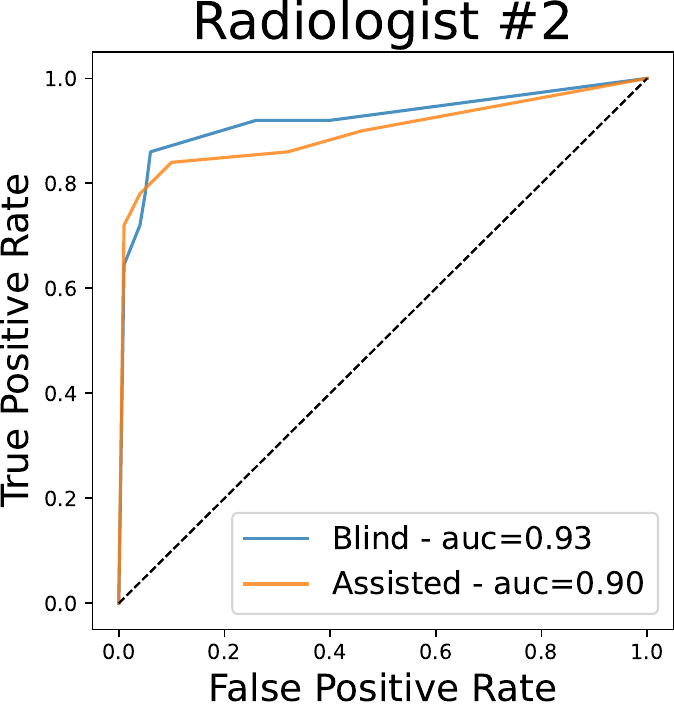}
        \includegraphics[width=0.16\linewidth]{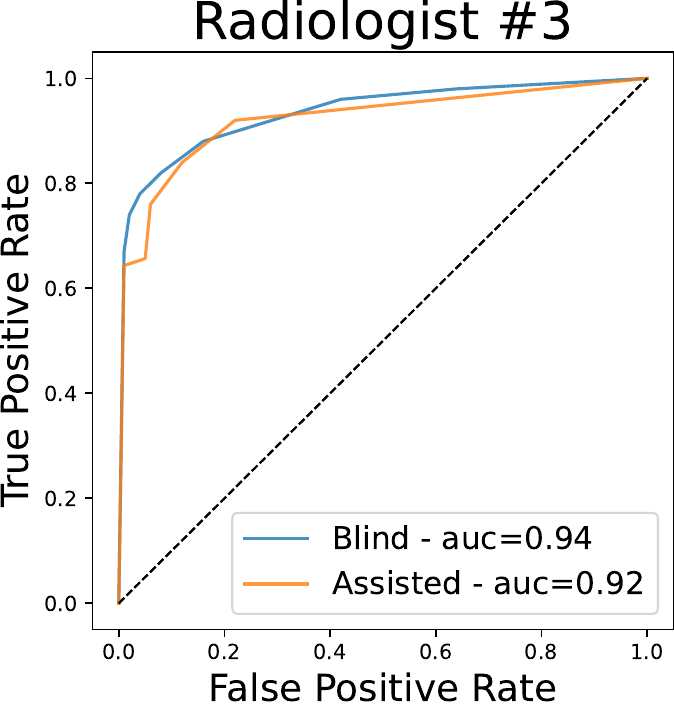}
        \includegraphics[width=0.16\linewidth]{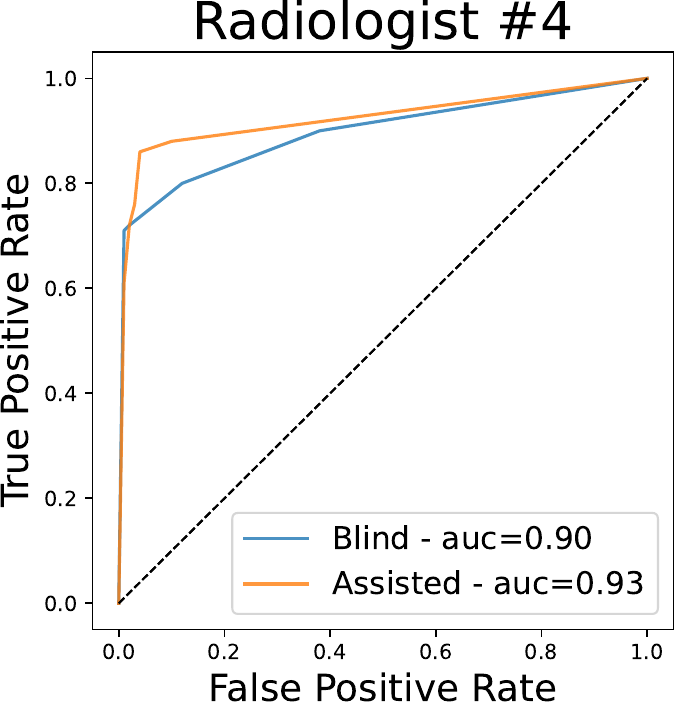}
        \includegraphics[width=0.16\linewidth]{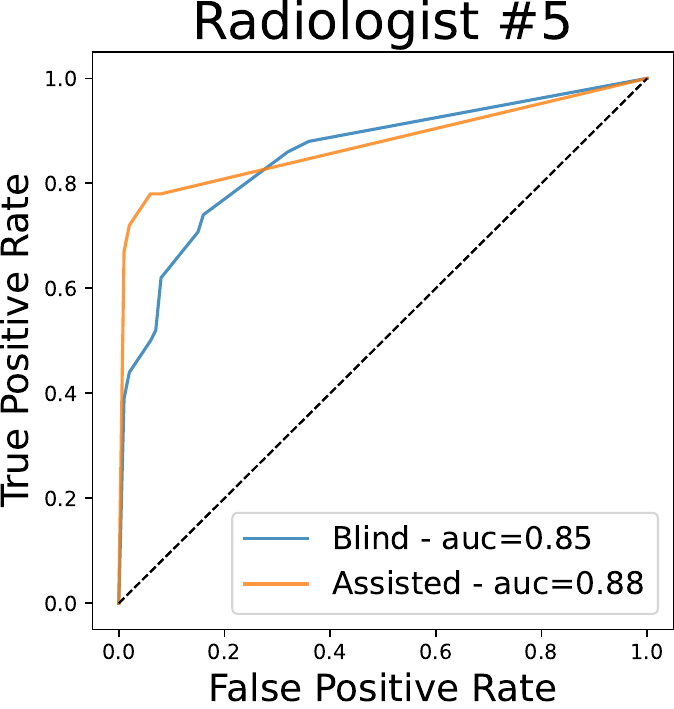}
        \includegraphics[width=0.16\linewidth]{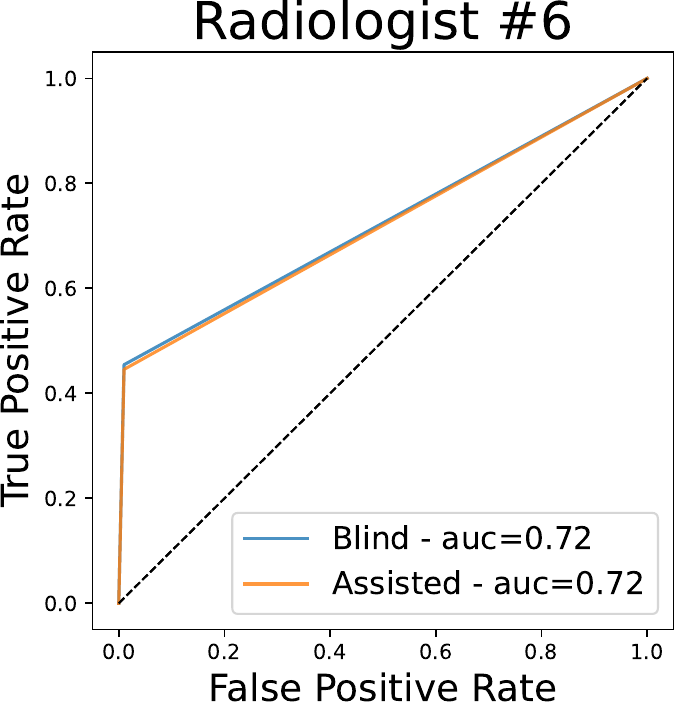}
      \caption{ROC-AUC}
    \end{subfigure}
    \begin{subfigure}[b]{\linewidth}
        \includegraphics[width=0.16\linewidth]{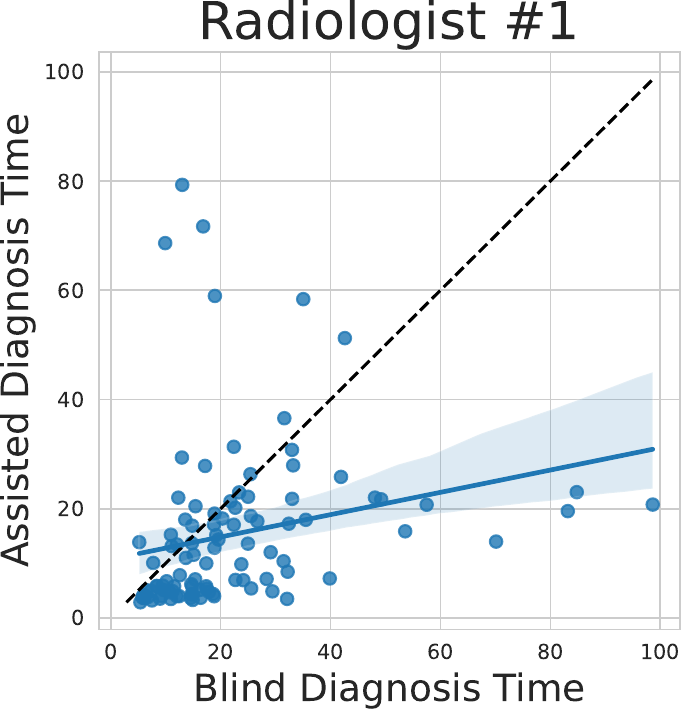}
        \includegraphics[width=0.16\linewidth]{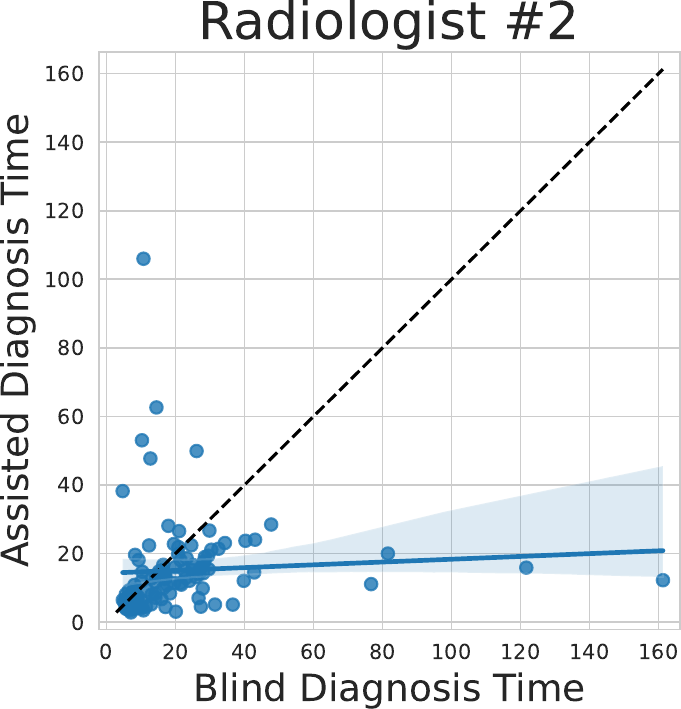}
        \includegraphics[width=0.16\linewidth]{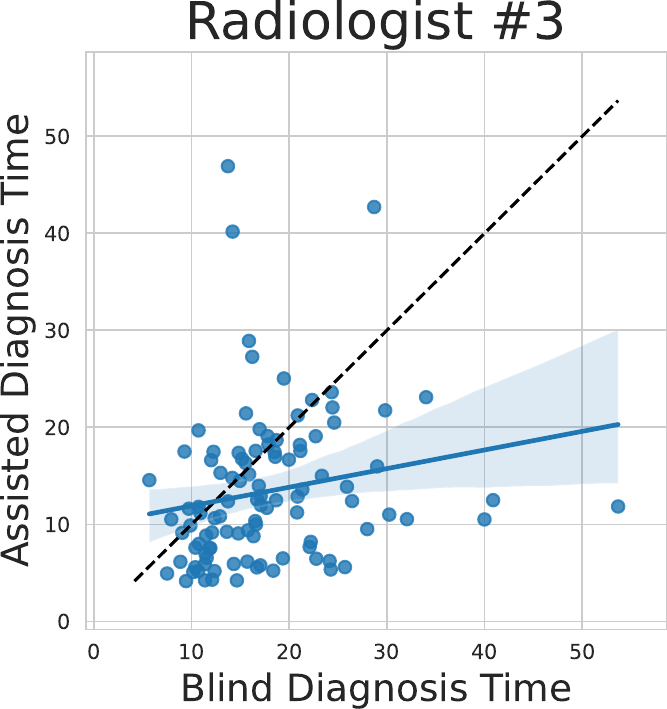}
        \includegraphics[width=0.16\linewidth]{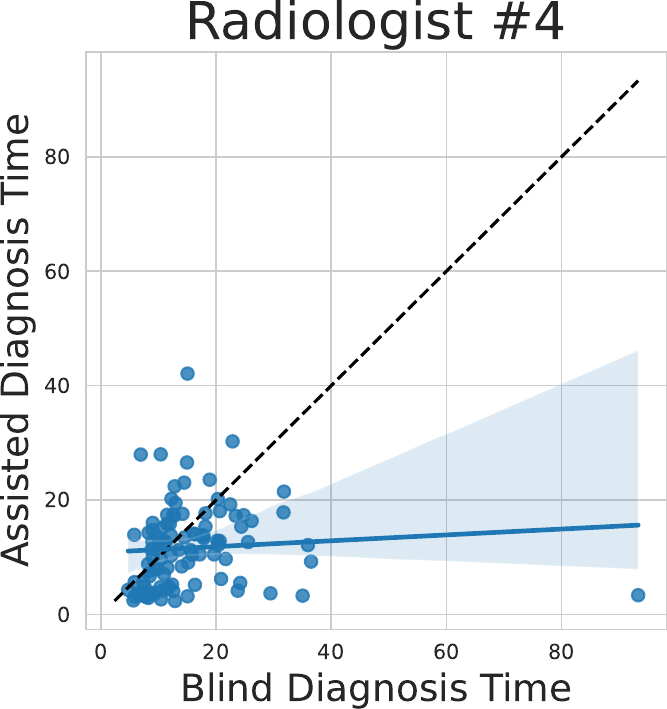}
        \includegraphics[width=0.16\linewidth]{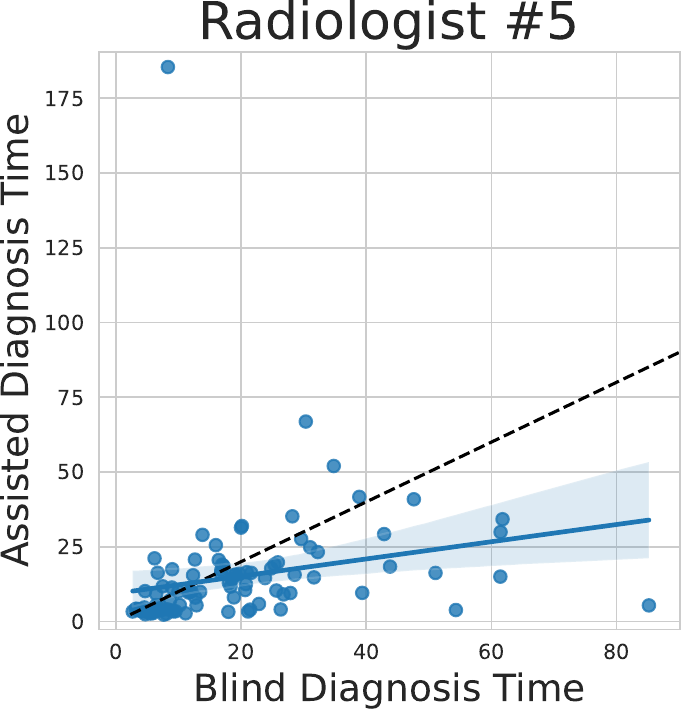}
        \includegraphics[width=0.16\linewidth]{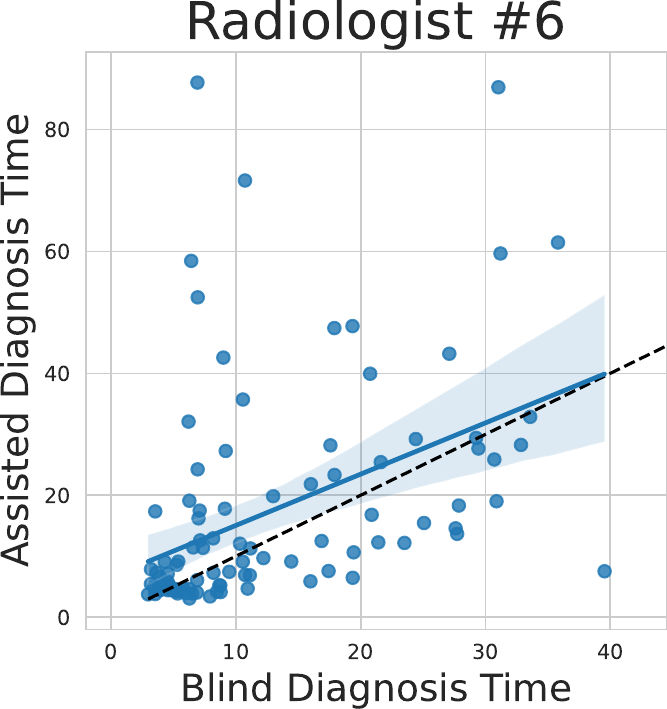}
      \caption{Diagnosis Time}
    \end{subfigure}
    \caption{Break-out of diagnosis performance by radiologist. (a) shows the AUC score of blind and AI-assisted evaluation. While for some radiologists, especially those with higher blind scores we notice a slight decrease in AUC, the improvement is consistent in those who achieve a lower base score. On average, the AUC increases from 0.85 to 0.88, as shown in Fig.\ref{fig:teaser}. (b) compares the blind diagnosis time with the AI-assisted diagnosis time. The blue line represents a linear regression line, and the dashed black line represents the identity. We can see that, on average, the diagnosis time decreases as the fitted line lies below the identity for almost all radiologists.}
    \label{fig:validation-breakout}
\end{figure*}

\subsection{Validation Protocol}

To perform the clinical evaluation, we employ a pool of 6 radiologists from Ospedale di Rivoli and ASL To3, with different experience levels. They are summarized in Tab.~\ref{tab:radiologists}.
For each image, a severity score~\cite{borghesi2020brixia} is assigned by each radiologist upon visual inspection, ranging from 0 (healthy) to 18 (maximum severity). During diagnosis, the radiologists do not know the correct label for a given patient. The experiment is repeated in two settings, with a wash-out period in between:
\begin{itemize}
    \item \emph{blind} setting: radiologists can only leverage the image to make a diagnosis
    \item \emph{AI-assisted} setting: the predicted probabilities of our DNNs are shown alongside the image.
\end{itemize}

\begin{table}[]
    \centering
    \begin{tabular}{c l c}
    \toprule
      \textbf{Initials} &  \textbf{Affiliation} & \textbf{Years of Experience} \\
    \midrule 
        S. N. & ASL To3 Rivoli & 1  \\ %
        V. N. & ASL To3 Rivoli & 1 \\  %
        D. C. & ASL To3 Rivoli & 2 \\  %
        M. B. & ASL To3 Rivoli & 5 \\  %
        F. L. & ASL To3 Rivoli & 10 \\ %
        P. M. & ASL To3 Rivoli & 15 \\ %
    \bottomrule
    \end{tabular}
    \caption{List of the radiologists involved in the study.}
    \label{tab:radiologists}
\end{table}

\subsection{Validation Results}

Fig.~\ref{fig:teaser} shows the overall results of the validation in terms of prediction AUC and diagnosis time. In Fig.~\ref{fig:mean-auc}, the AUC is computed on the radiologists' average severity score assigned for each patient. We observe an improvement in the assisted setting compared to the blind evaluation. In Fig.~\ref{fig:mean-time}, we plot the diagnosis time in the two settings: on average (blue line), the time required for diagnosis decreases in the assisted setting. 
In Fig.~\ref{fig:validation-breakout}, we show a breakout of the results for each participating radiologist for prediction AUC and diagnosis time. In most cases, we observe an improved performance in the assisted setting, with some significant improvement (e.g. AUC increases from 0.85 to 0.96 for radiologist \#1).
These results show that by employing AI systems in a real-world clinical setting, more accurate and faster diagnosis can be achieved.

\section{Conclusions}
\label{sec:conclusions}

In this work, we have presented the results from the Co.R.S.A. projects, which aimed at \emph{i.)} building an open and freely available data collection for Covid-19 diagnosis from CXR images \emph{ii.)} developing a robust deep-learning-based pipeline for automatic classification \emph{iii.)} performing a clinical validation in a real-world context, with a pool of expert radiologists. This project represents the results of our efforts in Covid-19 detection over the past few years. Furthermore, while the critical phase of the Covid-19 pandemic has passed, the foundation built by this project can serve as a robust basis for promptly initiating responses to future epidemics if necessary, thanks to the collaboration developed with different major hospitals and radiology units.
\newline

\noindent \textbf{Compliance with ethical standards}
This study was performed in line with the principles of
the Declaration of Helsinki. Approval was granted by 
Comitato Etico Interaziendale
A.O.U. San Luigi di Orbassano, AA.SS.LL. TO3,TO4,TO5 (10/24/2022/ No. 153/2022).
\newline

\noindent \textbf{Acknowledgments}
The project was funded through the ``INFRA-P2 - Potenziamento di laboratori di prova ed infrastrutture di ricerca già esistenti nella disponibilità di organismi di ricerca pubblici e progetti di ricerca e sviluppo finalizzati al contrasto della pandemia Covid-19'' call for proposals. A special acknowledgement goes to the other partners of the Co.R.S.A. project: Citt\`a della Scienza e della Salute and REGOLA s.r.l.. We wish to tank the valuable support of all the radiology units that contributed to the CORDA dataset.

\bibliographystyle{IEEEbib}
\bibliography{refs}

\begin{thebibliography}{10}

\bibitem{zu2020coronavirus}
Z.~Zu et~al.,
\newblock ``Coronavirus disease 2019 (covid-19): A perspective from china,''
\newblock {\em Radiology}, 2020.

\bibitem{shi2020radiological}
H.~Shi et~al.,
\newblock ``Radiological findings from 81 patients with covid-19 pneumonia in wuhan, china: a descriptive study,''
\newblock {\em The Lancet Infectious Diseases}, 2020.

\bibitem{corda_dataset}
M.~Alesina et~al.,
\newblock ``Corda dataset,'' \url{https://doi.org/10.5281/zenodo.7821611}, Apr. 2023.

\bibitem{glocker2019machine}
B.~Glocker et~al.,
\newblock ``Machine learning with multi-site imaging data: An empirical study on the impact of scanner effects,''
\newblock 2019.

\bibitem{tartaglione2020unveiling}
E.~Tartaglione et~al.,
\newblock ``Unveiling covid-19 from chest x-ray with deep learning: a hurdles race with small data,''
\newblock {\em IJERPH}, 2020.

\bibitem{barbano2022two}
C.~A. Barbano et~al.,
\newblock ``A two-step radiologist-like approach for covid-19 computer-aided diagnosis from chest x-ray images,''
\newblock in {\em ICIAP}, 2022.

\bibitem{irvin2019chexpert}
J.~Irvin et~al.,
\newblock ``Chexpert: A large chest radiograph dataset with uncertainty labels and expert comparison,''
\newblock in {\em AAAI}, 2019.

\bibitem{huang2017densely}
G.~Huang et~al.,
\newblock ``Densely connected convolutional networks,''
\newblock in {\em CVPR}, 2017.

\bibitem{barbano2023unbiased}
C.~A. Barbano et~al.,
\newblock ``Unbiased supervised contrastive learning,''
\newblock in {\em ICLR}, 2023.

\bibitem{borghesi2020brixia}
A.~Borghesi and R.~Maroldi,
\newblock ``Covid-19 outbreak in italy: experimental chest x-ray scoring system for quantifying and monitoring disease progression,''
\newblock {\em La Radiologia Medica}, 2020.

\end{thebibliography}

\end{document}